\documentclass{jfm}

\usepackage{graphicx} 
\usepackage[usenames]{color}

\usepackage{graphics}
\usepackage{epsfig}
\usepackage{eurosym}
\usepackage{amssymb}
\usepackage[english]{babel}

\newcommand{\be}{\begin{equation}}
\newcommand{\ee}{\end{equation}}
\newcommand{\bea}{\begin{eqnarray}}
\newcommand{\eea}{\end{eqnarray}}
\newcommand{\bfig}{\begin{figure}}
\newcommand{\efig}{\end{figure}}
\newcommand{\bc}{\begin{center}}
\newcommand{\ec}{\end{center}}
\newcommand{\btab}{\begin{tabular}}
\newcommand{\etab}{\end{tabular}}

\let\oldepsilon\epsilon
\let\epsilon\varepsilon
\let\varepsilon\oldepsilon

\newcommand{\Fr}{{\mathcal{F}}}

\newcommand{\Lsat}{L_{\rm sat}}

\title{The bar instability revisited}
\author[F. Chiodi, B. Andreotti, P. Claudin]
{F\ls I\ls L\ls I\ls P\ls P\ls O\ns C\ls H\ls I\ls O\ls D\ls I, \ns B\ls R\ls U\ls N\ls O\ns A\ls N\ls D\ls R\ls E\ls O\ls T\ls T\ls I\ns \and P\ls H\ls I\ls L\ls I\ls P\ls P\ls E \ns C\ls L\ls A\ls U\ls D\ls I\ls N} 
\affiliation{Laboratoire de Physique et M\'ecanique des Milieux H\'et\'erog\`enes,\\
(PMMH UMR 7636 ESPCI -- CNRS -- Univ. Paris Diderot -- Univ. P. M. Curie)\\
10 rue Vauquelin, 75231, Paris Cedex 05, France.}
\pubyear{???}
\volume{???}
\pagerange{??--??}
\date{\today}
\begin{document}
\maketitle

\begin{abstract}
The river bar instability is revisited, using a hydrodynamical model based on Reynolds averaged Navier-Stokes equations. The results are contrasted with the standard analysis based on shallow water Saint-Venant equations. We first show that the stability of both transverse modes (ripples) and of small wavelength inclined modes (bars) predicted by the Saint-Venant approach are artefacts of this hydrodynamical approximation. When using a more reliable hydrodynamical model, the dispersion relation does not present any maximum of the growth rate when the sediment transport is assumed to be locally saturated. The analysis therefore reveals the fundamental importance of the relaxation of sediment transport towards equilibrium as it it is responsible for the stabilisation of small wavelength modes. This dynamical mechanism is characterised by the saturation number, defined as  the ratio of the saturation length to the water depth $L_{\rm sat}/H$. This dimensionless number controls the transition from ripples (transverse patterns) at small $L_{\rm sat}/H$ to bars (inclined patterns) at large $L_{\rm sat}/H$.   At a given value of the saturation number, the instability presents a threshold and a convective-absolute transition, both controlled by the channel aspect ratio $\beta$. We have investigated the characteristics of the most unstable mode as a function of the main parameters, $L_{\rm sat}/H$, $\beta$ and of a subdominant parameter controlling the relative influence of drag and gravity on sediment transport. As previously found,  the transition from alternate bars to multiple bars is mostly controlled by the river aspect ratio $\beta$. By contrast, in the alternate bar regime (large $L_{\rm sat}/H$), the selected wavelength does not depend much on $\beta$ and approximately scales as $H^{2/3}L_{\rm sat}^{1/3}/C$, where $C$ is the Ch\'ezy number.
\end{abstract} 

\maketitle

\section{Introduction}

Bars are large scale river bedforms whose wavelength is comparable to channel width and whose height is comparable to flow depth. Due to these two large length-scales, bars interact with structures and navigation, and have a profound impact on several aspects of river engineering like river control (channel shift) and hazard prevention (bank failures). 

Free bars form in straight or weakly curved channels and  consist of repetitive sequences of migrating pools and diagonal depositional riffles. Alternate (single row) bars are the generic type of free bars in  sandy streams and gravel bed rivers, when the channel is narrow enough (Callander, 1969). When the channel presents a large width to depth aspect ratio (typically in glaciary valleys), the river develops multiple bars, leading to braiding patterns (Fujita \& Muramoto, 1985; Crosato \& Mosselman, 2009). Fluvial bars may also be forced by various effects like curvature, width variations or confluences. The formation of free alternate bars in an initially straight channel has been considered as the possible origin of incipient meandering (Fukuoka, 1989). The `bar theory' of river meandering implies the coexistence of free and forced bars in weakly curved channels and transition from migrating free to steady forced bars in developing meanders (Fujita and Muramoto, 1982; Tubino \& Seminara, 1990; Crosato et al., 2011).  This idea has been further developed theoretically in the framework of the so-called  'bend theory', in which the formation of meanders results from the non-linear interaction between self-excited (free) and curvature-driven (forced) bed responses (Seminara \& Tubino, 1989). Overall, the emergence of bars and their coupling with the shape of the banks are crucial ingredients required to understand the global morphodynamics of alluvial channels at the river scale (see the recent review of Seminara, 2010).

Since the seminal work of  Kennedy (1963), there is a wide agreement that bedforms --~both free bars and sand ripples~-- develop by linear instability of an erodible bed subject to a turbulent flow: infinitesimal perturbations of the bed (and therefore of the flow) are selectively amplified at certain wavelengths in the course of their downstream migration. By contrast with sand ripples, which are transverse to the flow, bars develop from a three-dimensional instability of the sediment bed  in which the most unstable modes are inclined with respect to the flow (Chang \& Simons, 1970; Chang et al., 1971; Schumm \& Khan, 1972;  Ikeda, 1983; Fujita \& Muramoto, 1985; Lisle et al., 1991, 1997; Lanzoni, 2000a, 2000b; Devauchelle et al., 2010a, 2010b; Andreotti et al., 2012). Like optical or acoustical wave-guides do, the channel selects a discrete number of guided unstable modes labelled by the number $m$ of rows of bar perturbations ($m = 1$ corresponds to alternate bars). The transverse wave-number is therefore directly selected by the channel width  (Parker, 1976). Several increasingly refined  linear stability analysis have been published in the literature (Callander, 1969; Engelund \& Skovgaard, 1973; Parker, 1976; Freds\o e 1978; Colombini et al., 1987; Garc\'\i a \& Ni\~no, 1993).  These theories predict the dispersion relation --~both the growth rate $\sigma(k_x)$ and the propagation velocity $c(k_x)$ of perturbations of wavenumber $k_x$~-- within the linear regime, the marginal stability conditions in the space of flow and sediment parameters and the wavenumber $k_{\rm max}$ selected by the instability mechanism, i.e. those corresponding to the maximum growth rate. The parameter controlling the instability is the width to depth ratio $\beta=W/H$ characterising the cross section of the channel. Colombini et al. (1987) have shown that nonlinear effects cause bed perturbations to reach an equilibrium finite amplitude characterized by periodic diagonal fronts. Under suitable conditions, which are rarely encountered in nature, the above equilibrium solution may in turn become unstable and bifurcate into a more complex quasi-periodic pattern (Schielen et al., 1993). One of the most important property of the linear instability is the existence of a particular value of $\beta$ for which the migration speed of the marginally stable mode vanishes (Blondeaux \& Seminara, 1985). This so-called `resonant' condition is the key ingredient of the `bend theory' of meandering, which assumes a weak coupling between bars and banks. River meanders are assumed to excite the bar instability without affecting the shape of the modes nor the dispersion relation; then, the coupling is optimal when non-growing free bars do not propagate with respect to the banks.

Recent reviews have claimed that the topic can be considered as fairly settled (Tubino et al., 1999; Seminara, 2010). However, most existing theories (all but Engelund \& Skovgaard (1973)) are based on depth-averaged Saint-Venant shallow water equations, which assume that the length-scale over which the flow varies is much smaller than the flow thickness $H$. Our primary goal, here, is to investigate the robustness of the results previously obtained, when the hydrodynamical disturbances are computed with Reynolds averaged Navier Stokes equations, which describe the vertical structure of the flow. Moreover, most of the previous papers do not take into account the fact that sediment transport adapts to a variation of the flow strength with a spatial lag. This sediment transport saturation transient, characterised by the so-called saturation length $L_{\rm sat}$, plays a dominant role in the linear instability of subaqueous sand ripples and aeolian dunes, as it controls the emergent wavelength (Andreotti et al., 2002; Lagr\'ee, 2003; Elbelrhiti et al., 2005; Valance, 2005; Valance \& Langlois, 2005; Claudin \& Andreotti, 2006; Charru, 2006; Fourri\`ere et al., 2010; Dur\'an et al., 2011). The saturation length allows one to build a new dimensionless number, the saturation number $L_{\rm sat}/H$, which compares the stabilising effect of the sediment transport relaxation transient to that of the free surface (Andreotti et al., 2012). Our second aim is to look at the influence of this dimensionless parameter on the bar instability.

In section~\ref{hydrodynamics}, we focus on three dimensional hydrodynamics above a modulated bottom, comparing the results obtained with Saint-Venant shallow water (SVSW) equations and with Reynolds averaged Navier-Stokes (RANS) equations. We show that shallow water approximation agrees semi-quantitatively with the full hydrodynamical resolution provided that the wavelengths in streamwise and perpendicular directions are larger than the flow thickness rescaled by the Ch\'ezy number. At small wavelength, however, they are quantitatively (wrong orders of magnitude) and qualitatively (wrong signs) incorrect. In section~\ref{1stADSL}, we investigate the consequences of these discrepancies on the bar linear stability analysis. We show that the maximum growth rate found in previous papers is an artefact of Saint-Venant equations which disappears when using three-dimensional hydrodynamics: large wavenumbers are incorrectly predicted to be stable. As a consequence, the introduction of the saturation length $L_{\rm sat}$ is necessary to recover an instability forming bars. We then show that the emergence of bedforms in a channel is controlled primarily by two parameters: the width to depth ratio $\beta=W/H$, as  found in former studies, but also the saturation length rescaled by the flow thickness $L_{\rm sat}/H$. In section~\ref{RANSonly}, we focus on the RANS based approach to revisit the basic properties of the bar instability, and investigate in particular the influence of the saturation parameter $L_{\rm sat}/H$ as well as the fluid shear velocity. We then study the convective-absolute transition of the instability. We finally present the stability diagram showing the transitions from ripples to alternate bars and from from alternate bars to multiple bars.
\begin{figure}
\centerline{\includegraphics{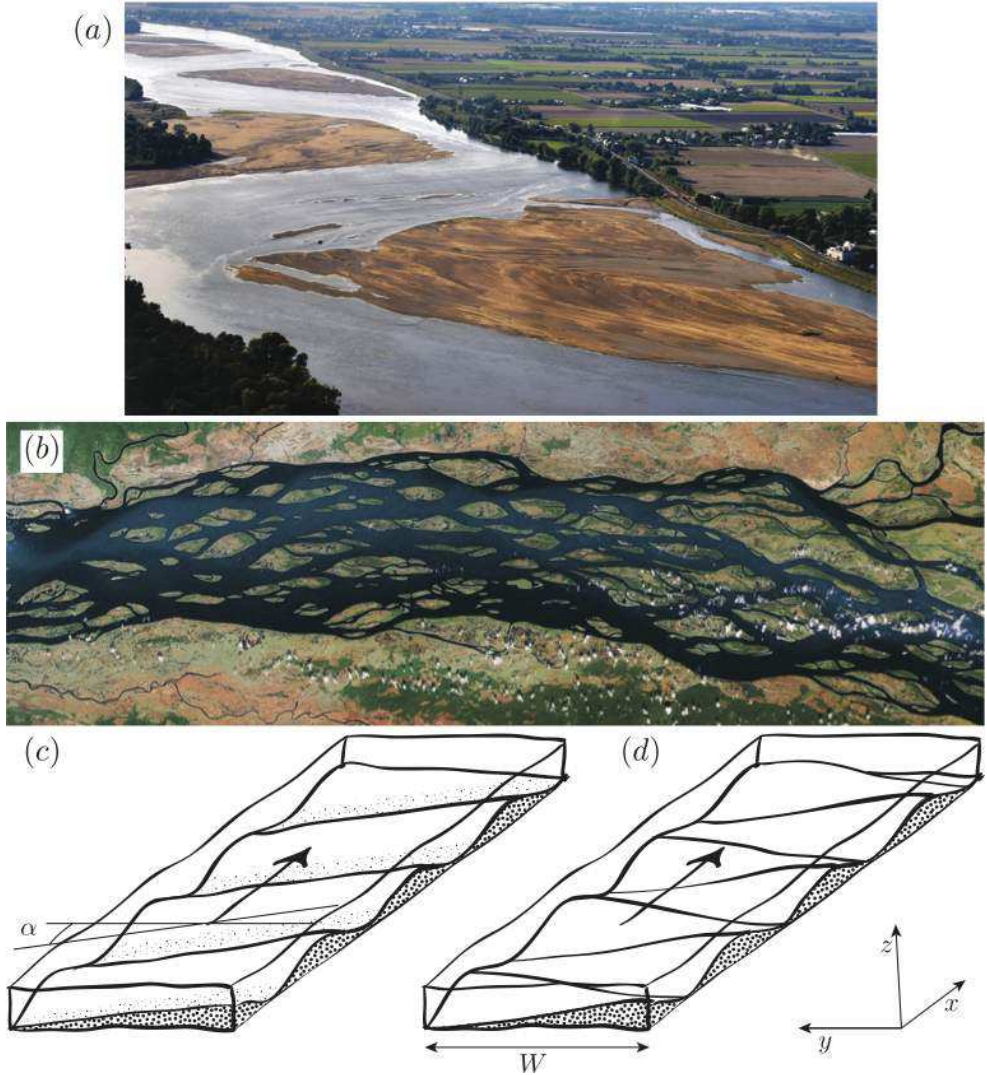}}
\caption{(a) Free alternate bars in the Loire river, close to St-Mathurin-sur-Loire (France; 47$^\circ$ 24' 50'' N, 0$^\circ$ 19' 04'' W). Photo credit: S. Rodrigues, universit\'e Fran\c cois Rabelais of Tours. These bars, of wavelength $\lambda \simeq 2000\,$m,  mostly evolve during annual floods, characterised by a typical flow depth $H\simeq 4\;$m. Sediment transport then mostly takes place in suspension, with a saturation length $\Lsat$ around $160\,$m. The saturation number $\Lsat/H$ is therefore around $40$. As the river width $W$ is around $480$~m, the aspect ratio $\beta=W/H$ is around $120$. (b) Multiple bars in the Congo river, with a similar mean wavelength around $\lambda \simeq 2000\,$m. The relevant flow depth $H$ is uneasy to define but is  smaller than $10\;$m. As the river width $W$ is around $9000$~m, the typical aspect ratio $\beta=W/H$ lies in the range $1000-2000$. (c) Schematic showing a plane wave mode inclined by an angle $\alpha$ with respect to the flow direction. (d) Schematic showing free alternate bars  in a channel of width $W$. They correspond to guided modes obtained by superimposing two plane waves propagating with opposite angles. The notations are the same as in Andreotti et al. (2012).}
\label{schematics}
\end{figure}
%

\section{Hydrodynamics over bedforms: contrasting Saint-Venant  approximation with a Reynolds averaged calculation}
\label{hydrodynamics}

In this section we focus on the purely hydrodynamical part of the problem. We introduce Saint-Venant shallow water and Reynolds averaged Navier-Stokes equations and compare their predictions for the modulation of the basal shear stress over a gently undulated bottom.

\subsection{Geometry and base flow}
\label{base_flow}
We consider a turbulent stream whose width $W$ is larger than the flow depth $H$. It flows on a plane inclined at an angle $\theta$ with respect to the horizontal. As schematised in Fig.~\ref{schematics}, $x$ is the direction of the flow, $y$ is in-plane transverse and $z$ is normal to the plane, oriented upwards. We do not describe the lateral boundary layers, of thickness $\sim H$ and assume that the banks  act as frictionless  impenetrable boundaries. Then, the base flow is similar to the homogeneous situation --~unbounded in the $x$ and $y$ directions~-- for which the pressure profile $p(z)$ is hydrostatic and the velocity profile $u_x(z)$ is well described by a logarithmic law (Julien, 1998):
\begin{eqnarray}
u_x(z) & = & \frac{u_*}{\kappa} \ln \left( 1 + \frac{z}{z_0} \right). \label{ubaseflow} \\
p(z) & = & \rho gH \cos\theta \left(1-\frac{z}{H}\right). \label{pbaseflow}
\end{eqnarray}
$z=0$ is the location of the bottom, where the velocity vanishes. $g$ is the gravity acceleration; $\rho$ is the fluid density, assumed constant; $u_* = \sqrt{gH\sin\theta}$ is the basal shear velocity;  $\kappa=0.4$ is the von K\'arm\'an phenomenological constant; $z_0$ is the hydrodynamical bottom roughness and is in general much smaller than $H$. Throughout this paper, we will use $H/z_0 = 10^2$, which is a typical value for natural rivers. The depth averaged velocity $\bar u = \frac{1}{H} \int_0^H u_x(z) dz$ is related to the shear velocity by the Ch\'ezy coefficient $C$:
\begin{equation}
C =\left(\frac{u_*}{\bar u}\right)^2\simeq \left(\frac{\kappa}{\ln \frac{H}{z_0}-1}\right)^2,
\label{Chezy}
\end{equation}
where we have used the hypothesis that $H/z_0 \gg 1$. The $y$- and $z$-components of the velocity are null. We define the Froude number $\Fr$ as the ratio of the surface velocity to the velocity of gravity surface waves in the shallow water approximation:
\begin{equation}
\Fr \equiv \frac{u_x(H)}{\sqrt{gH}} \simeq \frac{1}{\kappa} \sqrt{\sin \theta} \, \ln \frac{H}{z_0}.
\label{defFroude}
\end{equation}
This number can be of order unity in flumes but is in general small for large natural rivers, due to their small slopes.

\subsection{Linear response of a flow to steady bedforms}
\label{hydrodynamics-LineaResponse}
Before addressing the formation of bars, we discuss the linear response of hydrodynamics to bed perturbations. The technical aspects of the present analysis are detailed in Fourri\`ere et al. (2010) and Andreotti et al. (2012). Therefore, we shall here only sketch the main lines of the computations and refer the interested reader to these papers. As the base state of the bed is homogeneous in $x$ and $y$, the bedform elevation $Z(x,y)$ can be decomposed over normal Fourier modes. Without loss of generality, we can therefore consider one such mode
\begin{equation}
Z = -\tan\theta \, x + \zeta e^{ik(\cos\alpha \, x + \sin\alpha \, y)}=-\tan\theta \, x + \zeta e^{(ik_x\, x +ik_y \, y)}
\label{defZ}
\end{equation}
corresponding to undulations making an angle $\alpha$ with the direction of the flow (see Fig.~\ref{schematics}). $\vec k = (k_x,k_y)=(k \cos\alpha, k \sin\alpha)$ is the wave vector. We shall compute the first order linear response of the flow to this bed perturbation, describing hydrodynamics either by (i) depth averaged Saint-Venant equations, or (ii) Reynolds averaged Navier-Stokes equations.

In the SVSW approach, the governing equations read in the limit of small slopes
\begin{eqnarray}
\vec \nabla\cdot (h\vec u) & = & 0,
\label{StVenantcontinuity} \\
(\vec u\cdot \vec \nabla) \vec u & = & -g\vec \nabla (Z+h)-C\,\frac{|\vec u|\,\vec u}{h},
\label{StVenantmomentum}
\end{eqnarray}
where $\vec u = (u_x, u_y)$ is the two-dimensional depth averaged velocity and $h$ is the local water depth. In the homogeneous case, these equations have the simple solution $h=H$ and $\vec u=\sqrt{gH\tan\theta/C} \, {\vec e}_x$. In this approximation, the basal shear stress $\vec{\tau}^b$ is assumed to be a function of the velocity: $\vec{\tau}^b = -C \rho |\vec{u}| \vec{u}$. Importantly, the typical length-scale in the plane $(x,y)$ is $H/C$, which is much larger than the flow depth $H$ as a typical value of the Ch\'ezy number is $C \simeq 10^{-2}$.

The RANS equations have exactly the same form as Navier-Stokes equations,
\begin{eqnarray}
\partial_j u_j & = & 0, \label{NS1} \\
\partial_t  u_i + u_j \partial_j u_i & = & g_i- \frac{1}{\rho} \partial_i p -  \frac{1}{\rho} \partial_j \tau_{ij}. \label{NS2}
\end{eqnarray}
except that $\tau_{ij}$  is the Reynolds stress tensor, which includes both viscous and turbulent diffusion of momentum. Assuming a fully turbulent state, at large Reynolds number, we use here a first order Prandtl-like turbulent closure to relate $\tau_{ij}$ to the strain rate $\dot \gamma_{ij} = \partial_i u_j + \partial_j u_i$:
\begin{equation}
\tau_{ij} = \rho \kappa^2 L^2 |\dot \gamma| \left(\frac{1}{3} \chi^2 |\dot \gamma| \, \delta_{ij} - \dot \gamma_{ij} \right).
\label{tauijPrandtl}
\end{equation}
$\chi$ is another phenomenological constant in the range $2.5$--$3$. The mixing length is chosen equal to $L=(z+z_0)\sqrt{1-z/H}$, in order to recover the logarithmic flow profile for the base state (Eq.~\ref{ubaseflow}). By comparison to SVSW equations, there is one more parameter in RANS, the ratio $H/z_0$ associated to the vertical structure of the flow.
\begin{figure}
\centerline{\includegraphics{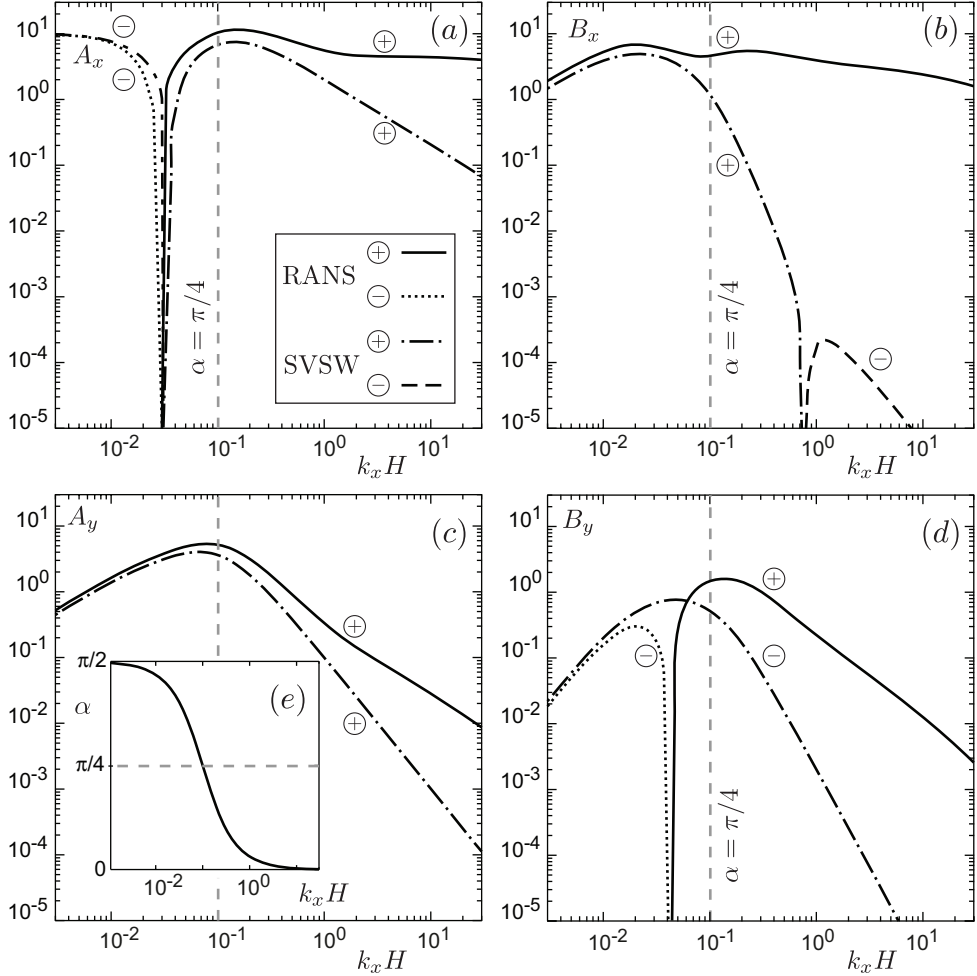}}
\caption{Discrepancies between hydrodynamical models at large $k_xH$. (a-d) Saint-Venant shallow water (SVSW) and Reynolds averaged Navier-Stokes (RANS) predictions of the  basal shear stresses coefficients $A_x$, $B_x$, $A_y$ and $B_y$. They have been computed for a fixed transverse wavenumber $k_y H=0.1$ and are plotted against $k_x H$. The other parameters are $\Fr = 0.1$ and $H/z_0 = 10^2$. Graphical convention for log-log plots: negative values are plotted in absolute value, but are displayed with a  type of line different from positive ones (see legend). (e) Angle $\alpha$ between $k_x$ and $k_y$ as a function of $k_x H$. In all panels, vertical grey dashed lines indicate the crossover from rather longitudinal patterns at small wavenumber $k_x H$ ($\alpha>\pi/4$) to rather transverse patterns at large $k_x H$ ($\alpha \le \pi/4$).}
\label{AxBxAyBy4kyH=0.1}
\end{figure}

The bed undulation amplitude $\zeta$ is assumed to be vanishingly small in front of the flow depth and the wavelength, so that all equations can be linearised with respect to the small parameter $k\zeta$. First order disturbances are computed for the pressure, for the water depth and for all the components of the velocity and of the stress tensor. As shown below, in the context of sediment transport and bedform evolution, the central quantity is the basal shear stress $\tau^b_{ij}$ (see section~\ref{1stADSL}). Denoting by a superscript $\wedge$ the space Fourier transform, the relationship of the basal stress linear response to the elevation profile $Z$ can be written as
\begin{equation}
{\hat \tau}^b_{xz} = - \rho u_*^2\,(A_{x} + iB_{x}) k \hat{Z},\quad {\rm and} \quad {\hat \tau}^b_{yz} = - \rho u_*^2\,(A_{y} + iB_{y}) k \hat{Z}.
\label{defAxBxAyBy}
\end{equation}
$A_x$, $B_x$, $A_y$ and $B_y$ are respectively the components of the basal shear stress in phase ($A_x$ and $A_y$) and in quadrature ($B_x$ and $B_y$) with the bed deformation. These coefficients have analytical expressions in the case of the Saint-Venant shallow water approximation, but are the result of numerical integration for the RANS equations. They are functions of $k_xH$ and $k_yH$, and depend on $\Fr$ and $H/z_0$. As developed in the next sections, the instability is controlled by the coefficients $B_x$ and $B_y$ and the bedform propagation speed by $A_x$ and $A_y$. One shall therefore pay attention to their sign and to their order of magnitude.
\begin{figure}
\centerline{\includegraphics{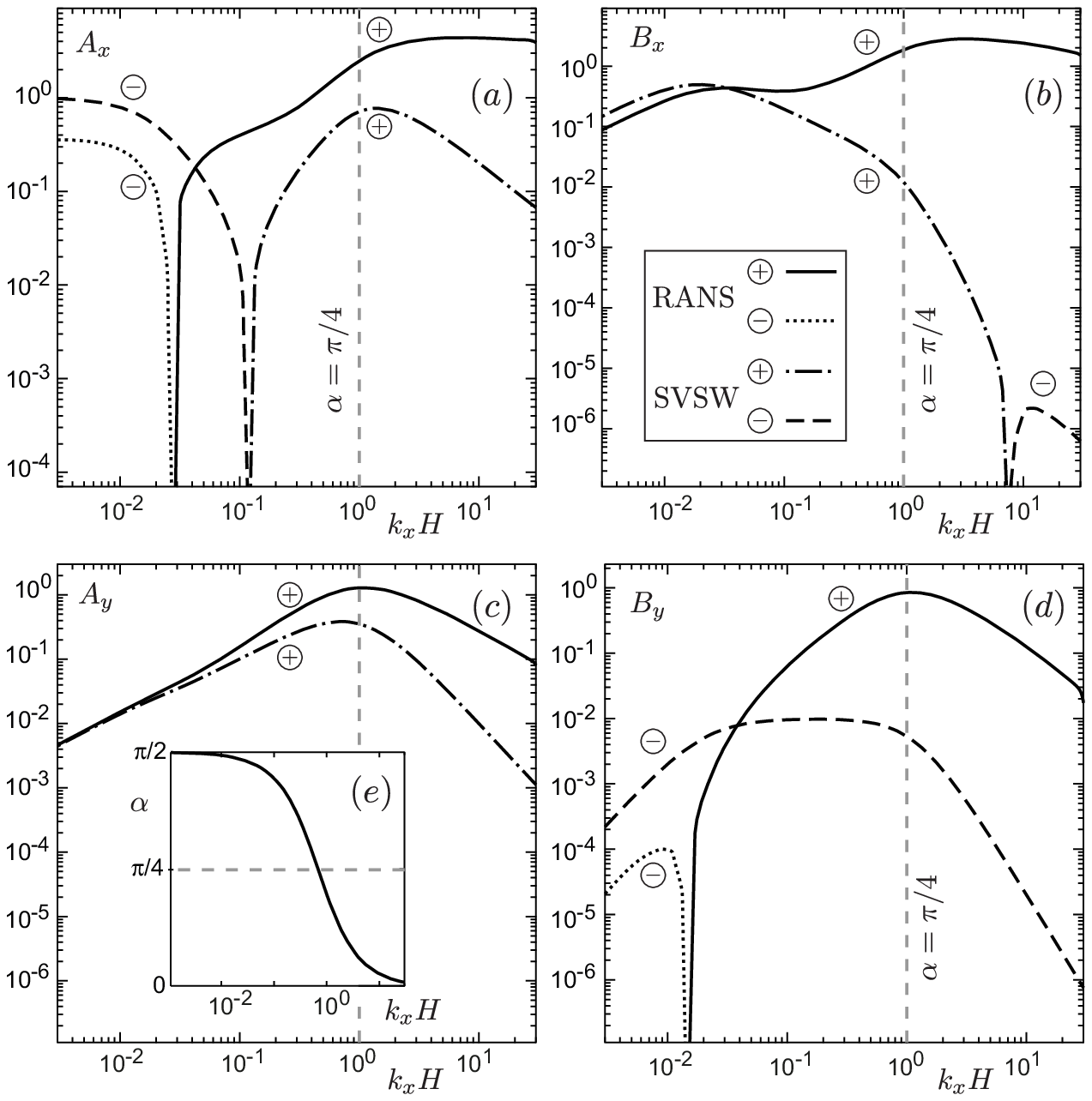}}
\caption{Discrepancies between hydrodynamical models at large $k_yH$. Same as Fig.~\ref{AxBxAyBy4kyH=0.1}, but for $k_y H=1$.}
\label{AxBxAyBy4kyH=1}
\end{figure}
%

\subsection{Basal shear stress coefficients}
\label{hydro:SVvsRANS}
As most former models of bar formation are based on SVSW equations, it is important to compare the behaviour of the basal shear stress coeffcients $A_x$, $B_x$, $A_y$ and $B_y$ obtained with this approximation, to the more reliable predictions of RANS. The stress coefficients are displayed in Figs.~\ref{AxBxAyBy4kyH=0.1} and \ref{AxBxAyBy4kyH=1} as a function of $k_x H$ for a fixed transverse wavenumber $k_y H$. The corresponding angle $\alpha = \arctan \frac{k_y}{k_x}$ is therefore continuously decreasing from $\pi/2$ in the limit $k_xH \to 0$ to $0$ in the limit $k_xH \to \infty$.

Let us first focus on the case of a small transverse wavenumber $k_y H = 0.1$ (Fig.~\ref{AxBxAyBy4kyH=0.1}). SVSW and RANS predictions are qualitatively similar for in-phase shear stress components: $A_x$ starts negative at small $k_xH$, changes sign and reaches a maximum around $k_xH=0.1$, while $A_y$ is always positive, also with a maximum around the same wavenumber. The scaling laws obtained with SVSW and RANS are however different at large $k_xH$.  The structure of the flow disturbance above bedforms does not resemble the base profile but presents a layered structure originally discussed in Jackson \& Hunt (1975). Such a structure cannot be described by the depth-averaged equations of the Saint-Venant approach. Moreover, the predictions for the components in quadrature are qualitatively different: RANS equations give an always positive $B_x$, while Saint-Venant equations predict that $B_x$ changes sign and become negative at large $k_xH$. RANS predicts, as observed experimentally (see the review Charru et al., 2013), that the basal shear stress is phase advanced with respect to topography while SVSW predicts a phase delay. As this phase between flow and relief is precisely at the origin of the emergence of bedforms, the consequences of this discrepancy are extremely important: the Saint-Venant approximation misses the instability at large wavenumber. For $B_y$, the situation is reversed: $B_y$ computed from the RANS equations switches from negative values at small wavenumber to positive ones, again with a maximum around $k_xH=0.1$, while it stays negative for all $k_xH$ in Saint-Venant modelling. This disagreement becomes worse for larger transverse wavenumber $k_y H = 1$ (Fig.~\ref{AxBxAyBy4kyH=1}), with an additional quantitative difference: the coefficient $A_x$ now changes sign at wavenumbers that differ by almost one decade from SVSW to RANS frameworks.

These differences are not surprising: contrarily to the lubrication approximation in the viscous regime, Saint-Venant equations are not well-controlled approximation of Navier-Stokes equations. As they rely on the shallow water approximation, one expects them to provide predictions semi-quantitatively valid  in the limit where $k_xH/C$ and $k_yH/C$ are both small. In the left sides of all panels of Fig.~\ref{AxBxAyBy4kyH=0.1}, when $k_xH$ is smaller than the Ch\'ezy number $C \simeq 10^{-2}$, the two hydrodynamical descriptions effectively match each other. However, Saint-Venant approximation leads to spurious results, including wrong signs for the most important shear stress coefficients, as soon as $k_x H/C$ or $k_y H/C$ are not small. Because we work here at fixed $k_y$ (see below), the SVSW equations become valid in the limit of inclined bedforms (large $\alpha$), when $k_yH$ is smaller than $C$. We emphasise that the prediction of RANS equations for the basal shear stress disturbance is very robust to the choice of the turbulent closure: although the flow in the outer layer can be sensitive to this choice, the structure of the flow in the inner layer remains almost unchanged  (Ayotte et al., 1994; Charru et al., 2013). Although experimental confirmations of the predictions are sparse (Poggi et al., 2007; Claudin et al., 2012), the RANS predictions can be considered as reliable in the turbulent regime.

\section{Consequences of hydrodynamical modelling for the bar instability}
\label{1stADSL}
We now wish to investigate the consequences of the differences between SVSW and RANS hydrodynamical predictions on the bar instability. In this section we therefore perform the linear stability analysis of a channelised flat erodible bed towards the formation of alternate bars.

\subsection{Sediment transport}
\label{sediment_transport}

We start from a generic description of sediment transport, characterised by a volumetric flux $\vec q = (q^x,q^y)$ defined as the volume of grains -- packed at the bed volume fraction -- passing per unit time trough a vertical surface of unit width and infinite vertical extension. The value $\vec q_{\rm sat}$ of this flux in the homogeneous steady state is said `saturated'. Both for bed load and suspended load, $\vec q_{\rm sat}$ is a function of the basal shear stress $\vec \tau^b= (\tau^b_{xz}, \tau^b_{yz})$ as well as of the longitudinal and transverse slopes of the bed. We write it under the generic following form:
\begin{equation}
\vec q_{\rm sat}=\Omega\,\left(\tau^b-\tau_{\rm th} \right)^\gamma \vec t,
\label{qsatgeneric}
\end{equation}
where $\Omega$ is a dimensionfull constant of proportionality, $\tau_{\rm th}$ is the threshold value below which transport vanishes, and $\gamma$ is a semi-empirical exponent: $\gamma \simeq 3/2$ for bed-load and turbulent suspension, see e.g. Meyer-Peter \& M\"uller (1948), Einstein (1950), Fernandez Luque \& van Beek (1976), Lajeunesse et al. (2010), Andreotti et al. (2012). The longitudinal slope of the bed modifies at the first order the threshold as
\begin{equation}
\tau_{\rm th} = \tau_{\rm th}^0 \left( 1 + \frac{1}{\mu} \partial_x Z \right),
\end{equation}
where $\mu \simeq 0.5$ is the avalanche slope (Dey, 2003; Fernandez Luque \& van Beek, 1976). Below we make use of the shear velocity threshold defined as $u^0_{\rm th} = \sqrt{\tau^0_{\rm th}/\rho}$. The transverse slope comes into the expression of the unit vector $\vec t$, which, at the first order, writes
\begin{equation}
\vec t=\vec e_\parallel - \frac{u^0_{\rm th}}{u_*}\,\frac{\partial_y Z}{\mu} \, \vec e_\perp,
\label{qsat2}
\end{equation}
where $\vec e_\parallel = \vec \tau^b /\tau^b$ is the unit vector in the direction of the basal shear stress and $\vec e_\perp = (-\tau^b_{yz},\tau^b_{xz})/\tau^b$ is its orthogonal counterpart (Andreotti et al. 2012). Combining the above expressions, the first order corrections to the saturation flux can be computed and read:
\begin{eqnarray}
\hat{q}_{\rm sat}^x & = & Q \left[  A_x + iB_x - \frac{i}{\mu} \left( \frac{u^0_{\rm th}}{u_*} \right)^2 \! \cos\alpha\right] k \hat{Z} \equiv Q (a_x+ib_x) k \hat{Z}, \label{qsatx1} \\
\hat{q}_{\rm sat}^y & = & \frac{Q}{\gamma} \left[ 1 - \left( \frac{u^0_{\rm th}}{u_*} \right)^2 \right] \left[ A_y+iB_y - \frac{i}{\mu} \frac{u^0_{\rm th}}{u_*} \, \sin\alpha \right] k \hat{Z} \equiv Q (a_y+ib_y) k \hat{Z}, \label{qsaty1}
\end{eqnarray}
where we have defined the reference flux $Q=\gamma \Omega \left[u_*^{2}-(u^0_{\rm th})^2 \right]^{\gamma-1} u_*^2$ and introduced the flux response coefficients $a_x$, $b_x$, $a_y$ and $b_y$. Note that transverse transport component vanishes at the sediment transport threshold, as $u_*\to u^0_{\rm th}$.

\subsection{Dispersion relation, assuming a locally saturated sediment transport}
\label{dispersion_relation}

In the case of an erodible bed, we assume a time scale separation between the dynamics of the flow and that of the bed, which is more slower. The flow can thus be considered as steady during the evolution of the bed profile,  governed by the sediment mass conservation equation:
\begin{equation}
\partial_t Z +\vec \nabla \cdot \vec q=0.
\label{conservemass}
\end{equation}
We first assume, as usually done in former articles on the formation of bars, that sediment transport is in local equilibrium with the flow, i.e. $\vec q= \vec{q}_{\rm sat}$ evaluated with the local basal shear stress. We will see that this strong hypothesis must be relaxed and that it is necessary to take the relaxation of transport towards saturation into account. 

This linear stability analysis is performed to study bar formation, i.e. in the geometry of a channelized flow, with straight banks located at $y =0$ and $y=W$. We assume that the base profiles (\ref{ubaseflow}, \ref{pbaseflow}) are valid in most of the channel. As already stated, we expect the influence of the banks to be limited to a boundary layer whose thickness is on the order of $H$. We do not describe this region close to the banks in details, and impose a slip boundary condition: the velocity of the fluid at the bank must be tangent to the bank itself. To obtain a mode of instability guided  by the channel, one superimposes plane waves of transverse wavenumbers $k_y$ and $-k_y$. The condition of null transverse velocity at the banks selects a discrete number of transverse wavenumbers labelled by the mode number $m$: $k_yW=m\pi$. Equivalently, this condition can be expressed in terms of the aspect ratio $\beta=W/H$ as:
\begin{equation}
k_yH=m \, \frac{\pi}{\beta} \, .
\label{betaky}
\end{equation}
$m=0$ corresponds to transverse bedforms (sand ripples) and $m = 1$ corresponds to alternate bars. We will first discuss the modes inclined with respect to the flow for a fixed order $m>0$. Then, we will summarise the results obtained for $m=0$ by Fourri\`ere et al. 2010. We will finally compare the growth rate of the different orders $m$ for the phase diagram of Fig.~\ref{diagramBetaLsatsH}. Defining the growth rate $\sigma$ and the phase velocity $c$ in $\hat Z(x,y,t) = \zeta \cos(k_y y) \exp [\sigma t + ik_x \, (x-ct)]$, one obtains, from Eq.~\ref{conservemass} and local equilibrium condition $\vec q= \vec{q}_{\rm sat}$, the following dispersion relation:
\begin{equation}
\sigma-ik_xc=-ik\,Q \left[k_x\,(a_{x} + ib_{x})+k_y\,(a_{y} + ib_{y})\right].
\label{DispRelLsat=0}
\end{equation}
As can be seen from Eqs.~\ref{qsatx1} and \ref{qsaty1} where the coefficients $a_x$, $b_x$, $a_y$ and $b_y$ are defined, both $\sigma$ and $c$ depend on the ratio $u_*/u_{\rm th}^0$ -- besides $k_xH$, $k_yH$, $\Fr$ and $H/z_0$.
\begin{figure}
\centerline{\includegraphics{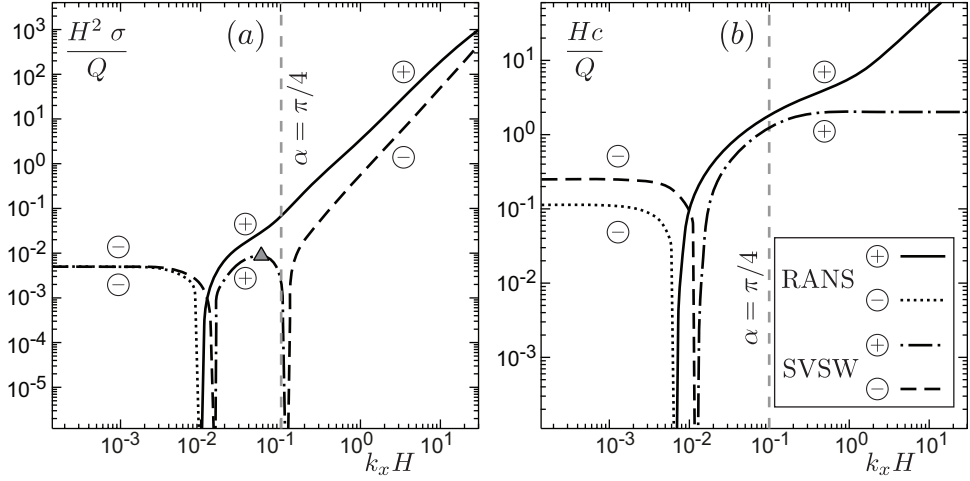}}
\caption{Consequences of the hydrodynamical model on the dispersion relation of the linear instability of a flat sediment bed. Rescaled growth rate $\sigma$ and phase velocity $c$ of inclined modes $m>0$ as functions of $k_xH$, computed with either RANS or Saint-Venant stress coefficients. Sediment transport is assumed to be locally saturated. These curves correspond to the limit $\Lsat=0$. These plots have been obtained for $\beta/m=31.4$ (i.e. for $k_y H=0.1$), $u_*/u_{\rm th}^0=2$, $\Fr = 0.1$ and $H/z_0 = 10^2$. The triangle indicates the location of the maximum growth rate on the Saint-Venant curve. The absence of such a maximum on the RANS curve shows that it is an artefact of SVSW approximation. The graphical convention is the same as in Fig.~\ref{AxBxAyBy4kyH=0.1} (see legend).}
\label{fig:RelDispLsatzero}
\end{figure}

The growth rate $\sigma$ and the phase velocity $c$ are plotted in Fig.~\ref{fig:RelDispLsatzero} as functions of $k_x H$, for a given set of parameters. One observes that the discrepancies between hydrodynamical predictions obtained with SVSW and RANS have fundamental consequences on the instability. A first difference affects the growth rate curves: while SVSW description shows an unstable region ($\sigma>0$) with a most unstable mode corresponding to an intermediate wavenumber $k_{\rm max} H \simeq 0.05$ (triangle in Fig.~\ref{fig:RelDispLsatzero}), the RANS hydrodynamical calculation does not present any maximum growth rate for $k_xH$ of order one or smaller. This fundamental discrepancy results from the fact that Saint-Venant equations predict the wrong sign for $B_x$ for this range of $k_xH$, i.e. a phase delay of the basal shear stress with respect to topography instead of a phase advance. Large wavenumbers thus appear to be artificially stabilised. Consequently, the selection of a maximum growth wavenumber determined using Saint-Venant equations should be taken as an artefact. This strong conclusion holds for all realistic values of $\beta$, $u_*/u_{\rm th}^0$, $\Fr$ and $H/z_0$. Although less crucial, the velocity curves also qualitatively differ in the large wavenumber limit, with asymptotic constant (SVSW) or growing (RANS) values of $c$.

The conclusion is twofold. First, the linear stability analysis previously proposed in the literature to explain the formation of bars suffer from the flaws of Saint-Venant shallow water, the observed maximum growth rate being an artefact. Second, with a more reliable hydrodynamical model, the growth rate does not present any maximum in the range of wavenumbers where bars are expected. We will now show that refining the sediment transport model by taking into account flux saturation transients allows the model to predict the formation of bars. We will see that SVSW and RANS based approaches can be partly reconciled when the so-called saturation length is introduced in the description.
\begin{figure}
\centerline{\includegraphics{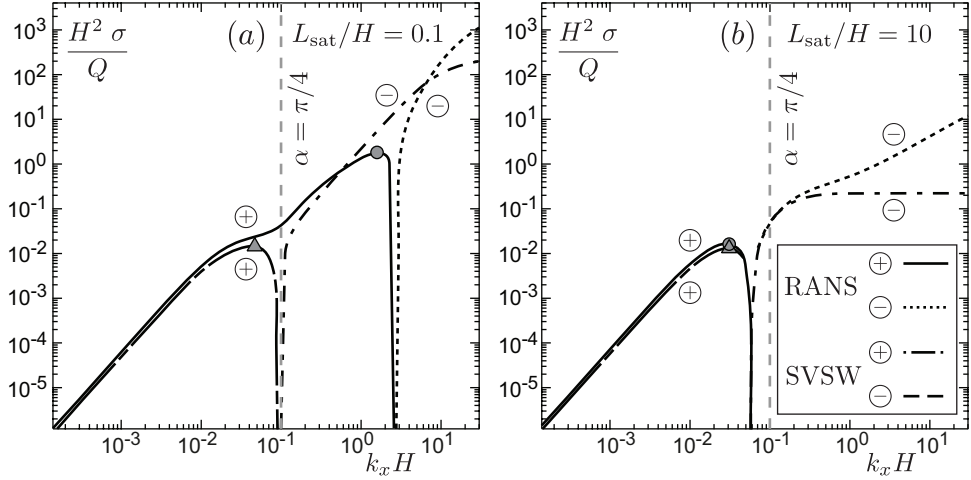}}
\caption{Comparison of the dispersion relation of inclined modes $m>0$ at small and large saturation parameter $L_{sat}/H$. Growth rate as a function of $k_x H$ for (a) $L_{sat}/H = 0.1$ and (b) $L_{sat}/H = 10$. The other parameters are $\beta/m = 31.4$ (i.e. for $k_y H=0.1$),  $u_* / u_{\rm th}^0 = 1$, $\Fr = 0.1$ and $H/z_0 = 10^2$. The triangle (resp. circle) indicates the location of the maximum growth rate on the Saint-Venant (resp. RANS) curve. The most unstable modes only coincide at large $L_{sat}/H$. At small $L_{sat}/H$, the maximum growth rate on the SVSW curve is due to a spurious description of hydrodynamics at large wavenumber. The graphical convention is the same as in Fig.~\ref{AxBxAyBy4kyH=0.1} (see legend).}
\label{fig:lsat1}
\end{figure}
%

\subsection{Modified dispersion relation, introducing the saturation length}
\label{Lsat}

In non-homogeneous situations, the sediment flux $\vec q$ does not equilibrate immediately with the local values of the basal shear stress, but relaxes towards its saturated value $\vec q_{\rm sat}$ over a typical distance called the saturation length $L_{\rm sat}$. We describe this process by the following first order relaxation equation:
\begin{equation}
L_{\rm sat} \left(\vec t \cdot \vec \nabla\right) \vec q=\vec q_{\rm sat}-\vec q,
\label{sateq}
\end{equation}
This saturation equation can both describe bed load (Charru 2006, Dur\'an et al. 2012, Charru et al. 2013) and suspended sediment transport (Claudin et al. 2011). In the later case, which is the most important for fluvial bars, the description has been validated using experimental data. The saturation length $L_{\rm sat}$ is an increasing function of the shear velocity $u_*$ and importantly presents a sharp increase at the transition from bed-load to suspended sediment transport. The saturation length is effectively set by the grain diameter $d$ for bed-load (the order of magnitude is around $10~d$) but rather set by the flow depth $H$ for a turbulent suspension. As an example, for the Loire river shown in Fig.~\ref{schematics}a, the saturation length increases by four orders of magnitude during floods, when the transition from bed load to suspended load is crossed. 

The description of the saturation transient brings a new dimensionless number in the problem, that we propose to call the saturation number, defined as the ratio $\Lsat/H$ between the saturation length and the flow depth (Andreotti et al., 2012). By construction, the saturation number is larger than $1$ for suspended load: it is typically on the order of $10^1$ or $10^2$ during flooding events. It can also be larger than $1$ in braided gravel streams where the particles can have a size comparable to the flow thickness. 
\begin{figure}
\centerline{\includegraphics{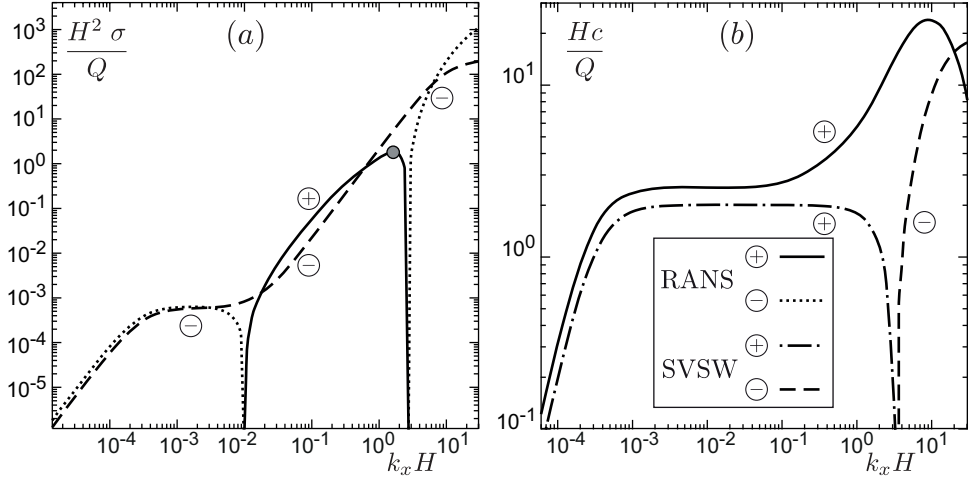}}
\caption{Consequences of the hydrodynamical model on the dispersion relation of transverse modes $m=0$. Rescaled growth rate $\sigma$ and phase velocity $c$ as functions of $k_xH$, computed with either RANS or Saint-Venant stress coefficients. These plots have been obtained for $L_{sat}/H = 0.1$,  $u_* / u_{\rm th}^0 = 1$, $\Fr = 0.1$ and $H/z_0 = 10^2$. The circle indicates the location of the maximum growth rate on the RANS curve. In Saint-Venant approximation the instability of transverse modes is missed ($\sigma<0$). The graphical convention is the same as in Fig.~\ref{AxBxAyBy4kyH=0.1} (see legend).}
\label{fig:mzer}
\end{figure}

The saturation transient equation gives in Fourier space
\begin{equation}
\hat{q}^i = \frac{1}{1+ik_x L_{\rm sat}} \, \hat{q}^i_{\rm sat}.
\label{q1}
\end{equation}
The dispersion relation then becomes
\begin{equation}
\sigma-ik_xc=-\frac{ik\,Q}{1+ik_x L_{\rm sat}} \left[k_x\,(a_{x} + ib_{x})+k_y\,(a_{y} + ib_{y})\right].
\label{DispRel}
\end{equation}
%
The new factor $(1+ik_x L_{\rm sat})^{-1}$ shows that the transport saturation transient has a stabilising effect on wavelengths comparable or smaller than $\Lsat$. The dispersion relation is plotted in Fig.~\ref{fig:lsat1} for two values of $\Lsat/H$. On all curves, the growth rate $\sigma$ is now negative at large wavenumber and presents a maximum controlled by $L_{\rm sat}$. Comparing SVSW and RANS based approaches, the most unstable mode is located at different values of $k_x H$ for small $\Lsat/H$, but coincides at large $\Lsat/H$. When the saturation length is large enough, the transport saturation transient stabilises all wavenumbers, for which Saint-Venant equations are not reliable. The two predictions then coincide in the whole range of unstable wavenumbers.  

To conclude this paragraph, we now consider the mode $m=0$, which corresponds to bedforms transverse to the flow. Fig.~\ref{fig:mzer} compares the dispersion relation obtained for SVSW and RANS hydrodynamical models, for a small but finite value of $\Lsat/H$.  As already noted by different authors, the Saint-Venant completely misses the instability of these modes, and therefore the emergence of ripples. By contrast, using a more reliable hydrodynamical calculation, one observes that transverse bedforms are unstable in a range  wave-numbers going from a lower cut-off scaling on $1/H$ to an upper cut-off scaling on $1/L_{\rm sat}$. We therefore reach a new important conclusion regarding the instability of a flat sediment bed: ripples ($m=0$) and bars ($m>0$) form by the very same instability, but possibly in different regimes. In order to find the most unstable mode, one must compare all the modes, including $m=0$. Due to the artefacts of Saint-Venant approach at large $k_xH$, this mode has been ignored so far in the discussion of the bar instability. 
\begin{figure}
\centerline{\includegraphics{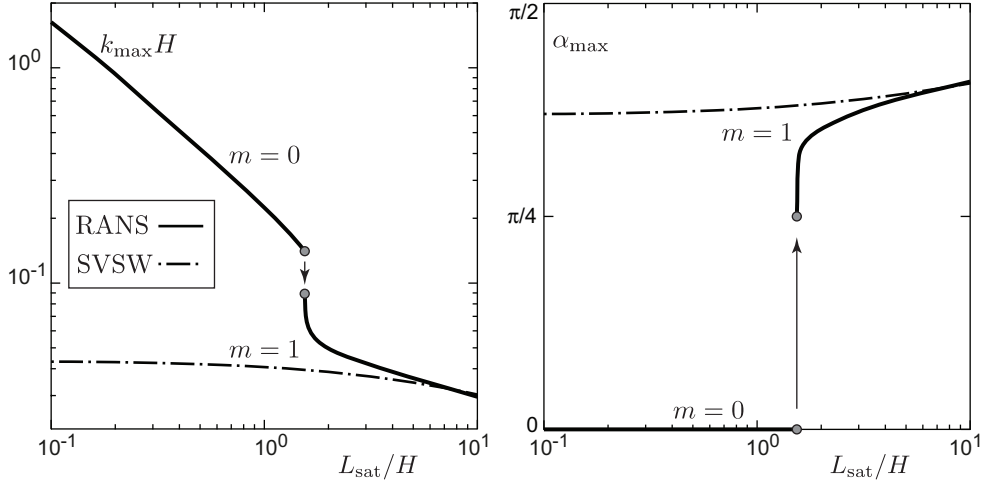}}
\caption{Transition from transverse ($m=0$) to inclined ($m=1$) bedforms controlled by the saturation number $\Lsat/H$. Longitudinal wavenumber $k_{\rm max}$ (a) and angle $\alpha_{\rm max}$ (b) of the most unstable mode as a function of $\Lsat/H$. The other parameters are $\beta = 31.4$ (i.e. $k_y H=0.1$ for $m=1$), $u_* / u_{\rm th}^0 = 1$, $\Fr = 0.1$ and $H/z_0 = 10^2$. At the transition point (circles), the maximum growth rates of modes $m=0$ and $m=1$ are equal.}
\label{fig:lsat2}
\end{figure}
%

\section{Revisiting the bar instability}
\label{RANSonly}

In the first part of the article, we have shown that the maximum growth rate deduced from Saint-Venant equations, when the transport saturation transient is ignored, is a pure arte-fact. We have then shown that the introduction of the saturation length is necessary to address the bar instability, once hydrodynamics is properly described. We will now focus on the RANS based approach to revisit the basic properties of the instability, and produce stability and bedform-phase diagrams. When necessary, we will still show the differences with the SVSW based approach.

\subsection{Influence of the saturation parameter $\Lsat/H$}
\label{influence_LsatsurH}
In order to study in more details the influence of the value of $\Lsat/H$ on the maximum growth rate, we have systematically varied this ratio between the two values used in Fig.~\ref{fig:lsat1}. We display in Fig.~\ref{fig:lsat2} the longitudinal wavenumber $k_{\rm max}$ and the wave angle $\alpha_{\rm max}$ of the most unstable mode. One observes that the saturation parameter $\Lsat/H$ triggers a sharp transition from transverse ($m=0$) to inclined bedforms ($\alpha_{\rm max}$ switches from $0$ to $\simeq 3\pi/8$). At the transition, located at $\Lsat/H \simeq 1.6$ for the parameters chosen to compute Fig.~\ref{fig:lsat2}, the wavenumber $k_{\rm max}$ drops by half a decade. Below the transition, one observes in Fig.~\ref{fig:lsat2}a that $k_{\rm max} H$ decreases as the inverse of the saturation parameter. Therefore, the wavelength of the transverse patterns ($\alpha_{\rm max}=0$) is proportional to $\Lsat$, which corresponds to sand ripples. Beyond the transition, the selected mode corresponds to alternate bars ($\alpha_{\rm max}>0$). The most amplified wavenumber presents a much weaker dependance on $\Lsat$. The wavelength approximately scales as $H^{2/3}L_{\rm sat}^{1/3}/C$. Importantly, the SVSW approach completely misses this transition from transverse to inclined bedforms. 

\subsection{Influence of the ratio $u_*/u_{\rm th}^0$}
\label{influence_usuruth}

\begin{figure}
\centerline{\includegraphics{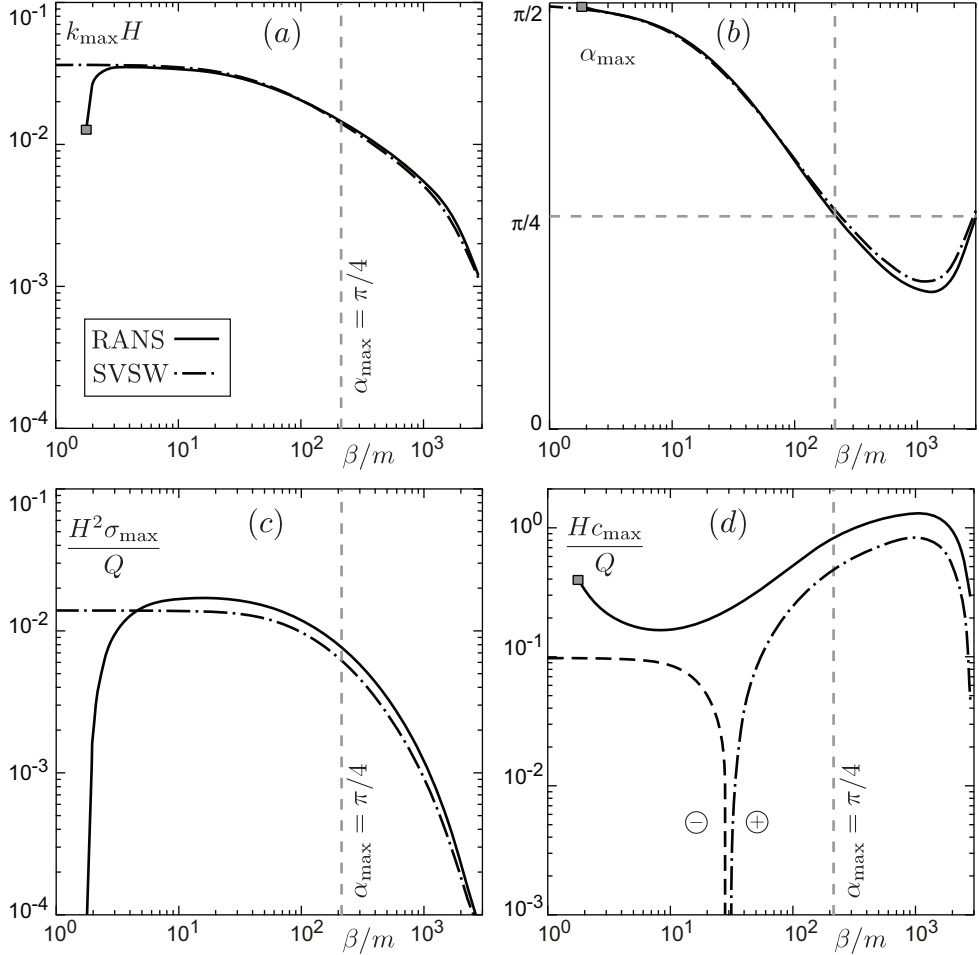}}
\caption{Rescaled horizontal wavenumber (a), wave angle (b), growth rate (c) and phase velocity (d), all computed for the most unstable mode, as functions of the channel aspect ratio divided by the mode number. These curves have been computed for $u_*/u_{\rm th}^0=1$. The other parameters are $\Lsat / H = 10$, $\Fr = 0.1$ and $H/z_0 = 10^2$. The graphical convention is the same as in Fig.~\ref{AxBxAyBy4kyH=0.1} (see legend).}
\label{fig:usuruth=1}
\end{figure}

\begin{figure}
\centerline{\includegraphics{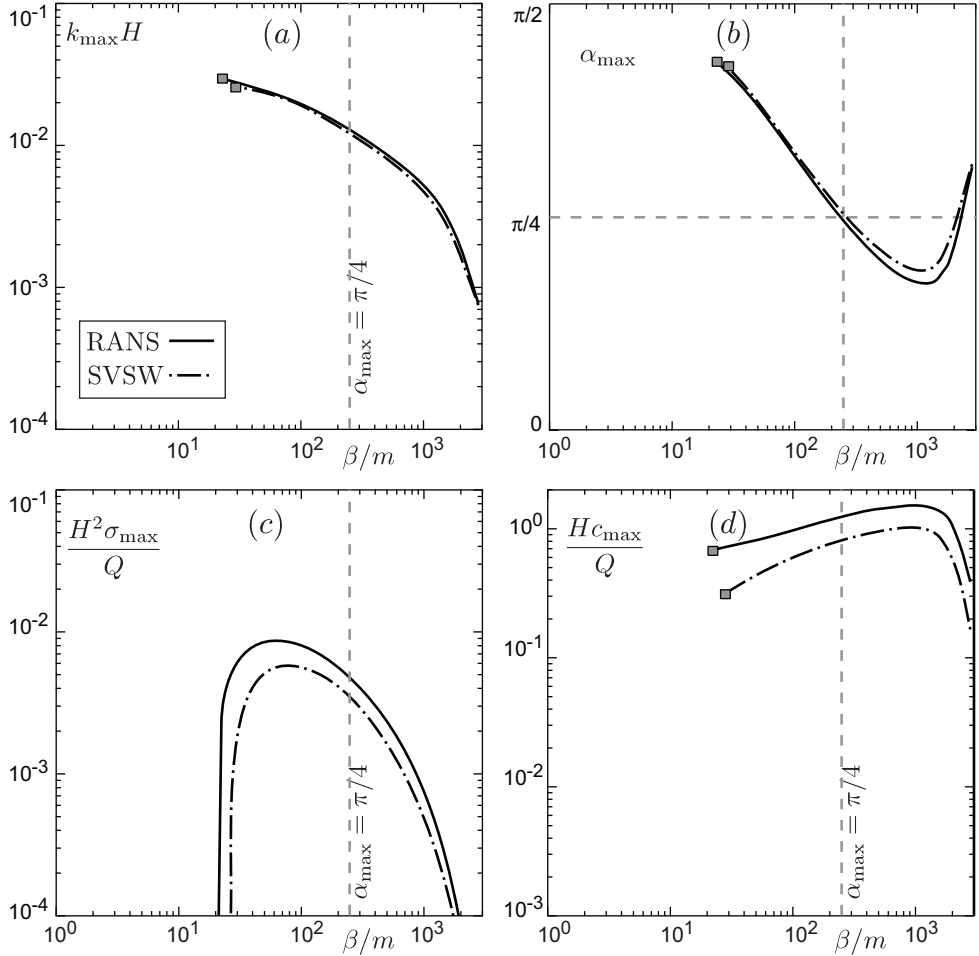}}
\caption{Same as Fig.~\ref{fig:usuruth=1}, but for $u_*/u_{\rm th}^0=2$.}
\label{fig:usuruth=2}
\end{figure}

\begin{figure}
\centerline{\includegraphics{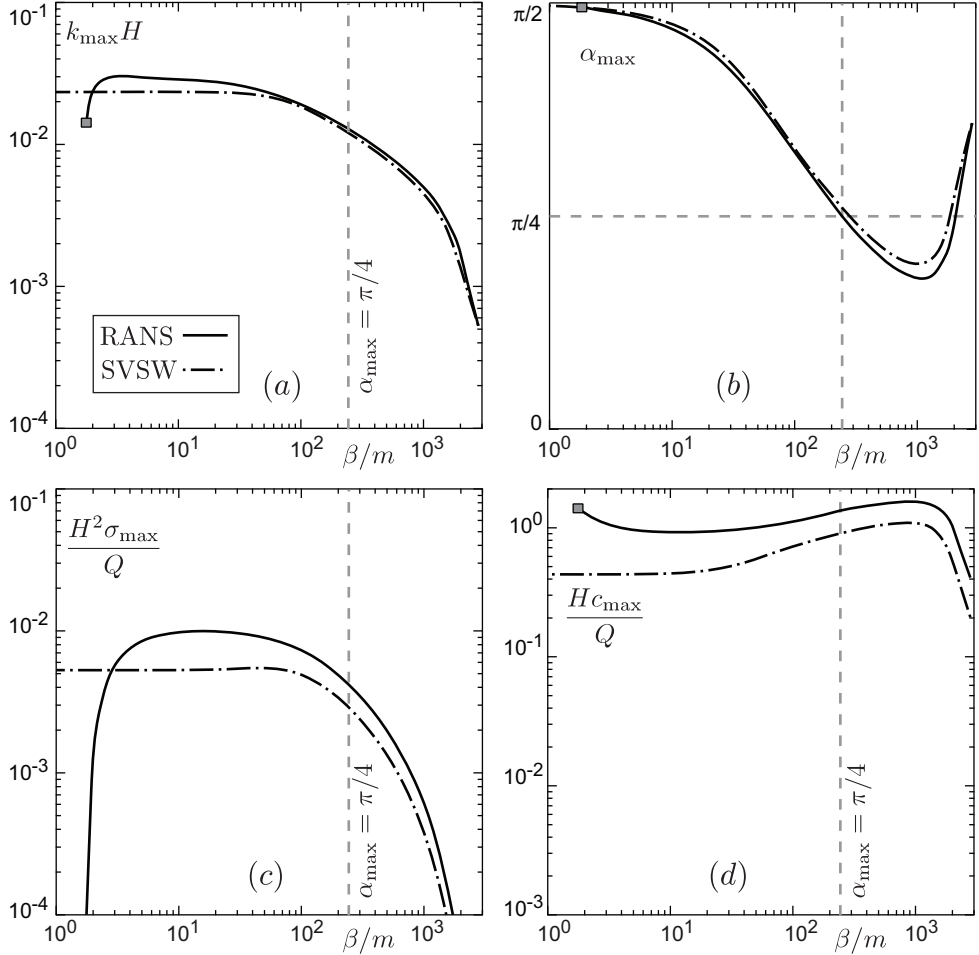}}
\caption{Same as Fig.~\ref{fig:usuruth=1}, but for $u_*/u_{\rm th}^0 \to \infty$.}
\label{fig:usuruth=infty}
\end{figure}

In this subsection we study the influence of the ratio $u_*/u_{\rm th}^0$ on the dispersion relation (\ref{DispRel}). We focus on the large saturation number regime, for which bars ($m>0$) are more unstable than ripples ($m=0$). This ratio enters the problem trough the flux coefficients $a_x$, $b_x$, $a_y$ and $b_y$ (see Eqs. \ref{qsatx1}, \ref{qsaty1}), when the influence of the longitudinal and transverse slopes on the shear threshold is taken into account. In Figs.~\ref{fig:usuruth=1}-\ref{fig:usuruth=infty}, we plot the longitudinal wavenumber $k_{\rm max}$, the wave angle $\alpha_{\rm max}$, the growth rate $\sigma_{\rm max}$ and the phase velocity $c_{\rm max}$ of the most unstable mode as functions of the channel aspect ratio $\beta$, for three different values of $u_*/u_{\rm th}^0$. The saturation length is fixed at the value $\Lsat/H=10$, for which we have seen above that Saint-Venant and RANS based predictions are comparable.

Close to the transport threshold, for $u_*/u_{\rm th}^0 \to 1$, there is no influence of the transverse sediment transport ($a_y \to 0$ and $b_y \to 0$). There is a stabilising effect of the longitudinal slope which vanishes as $\alpha$ tends to $\pi/2$. In the other limit, when $u_*/u_{\rm th}^0 \to \infty$, the transverse sediment transport becomes important, but the slope effect disappears. In the intermediate range of shear velocities, both transverse sediment transport and slope effects are present. In particular, the transverse slope effect has a stabilising effect when $\alpha$ tends to $\pi/2$. Looking only at the results based on the RANS approach, the general shape of the curves is very similar. For a given mode $m>0$, the bar instability presents a threshold controlled by $\beta/m$. In both limits of shear velocities close to the transport threshold and far from it, this instability threshold is mostly due to a decrease of the stress component $B_x$ at large $k_yH$ (we recall that $B_x$ controls the destabilising effect). In the intermediate range of shear velocities, the transverse slope effect adds a further stabilising effect which leads to an increased value of the threshold aspect ratio, $\beta_c$. Above this threshold, one can distinguish three zones of $\beta/m$. At small $\beta/m$, the wavelength is controlled by $H/C$, the angle is close to $\pi/2$, and the propagation speed increases with $\beta$. In the intermediate range of $\beta/m$, the angle decreases with $\beta$; the wavelength roughly scales as $\sqrt{HW/C}$, and the propagation speed still increases with $\beta$. Finally, at high $\beta/m$, the growth rate drops. For a given $\beta$, modes of higher $m$ are therefore more amplified. 

Comparing the results obtained with SVSW equations to those obtained with RANS approach, one observes a global agreement in the trends. However, for $u_*/u_{\rm th}^0 \to 1$ and $u_*/u_{\rm th}^0 \to \infty$, Saint-Venant equations miss the instability threshold at small $\beta$, i.e. when $k_yH/C$ is too large to get correct hydrodynamical results. Furthermore, these equations predict a change of sign of the propagation speed $c_{\rm max}$, close to the transport threshold, which is due to a wrong sign of $A_x$ at large transverse wavenumbers, see Fig.~\ref{AxBxAyBy4kyH=1}a. Interestingly, the results obtained with both models are very close to each other in the intermediate range of shear velocities, thanks to the stabilising transverse slope effects:  as the threshold $\beta_c$ is higher, Saint-Venant equations are correct in the whole range of unstable modes.
\begin{figure}
\centerline{\includegraphics{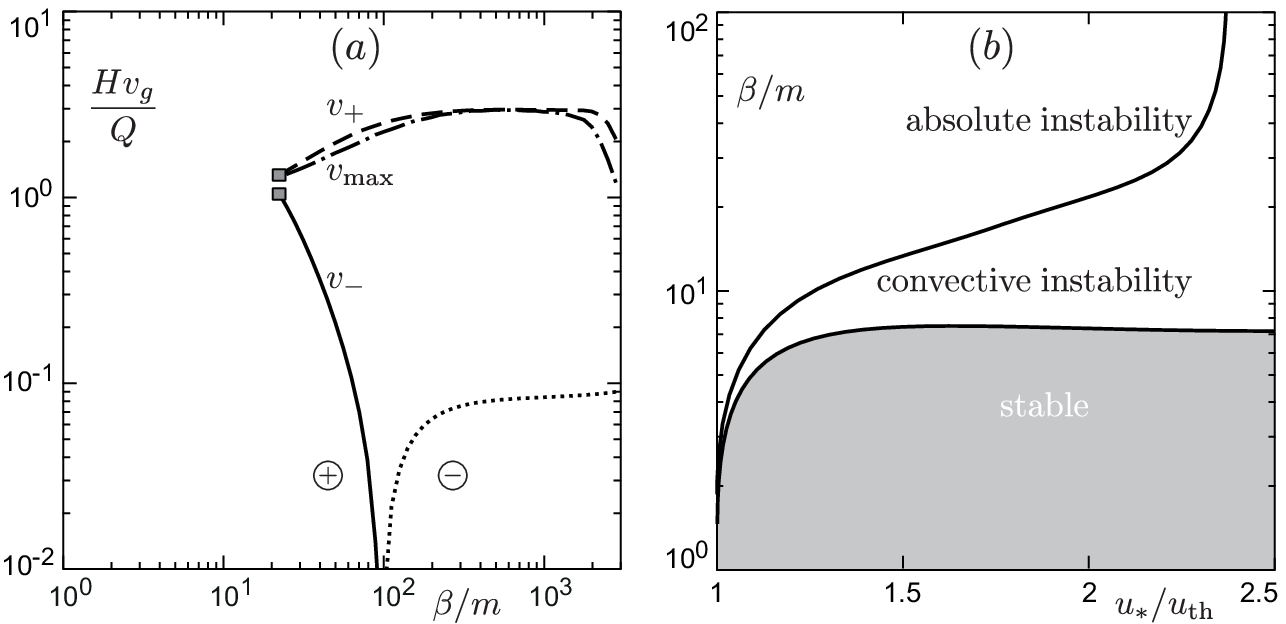}}
\caption{(a) Group velocity $v_g$ as a function of $\beta/m$. The three curves correspond to three particular points of the dispersion relation $\sigma(k_x H)$: the most unstable mode (dotted dashed line), and the two marginal modes at high (dashed line) and low (solid and dotted lines) wavenumbers. The latter changes sign, and we use a graphical convention similar to that described in Fig.~\ref{AxBxAyBy4kyH=0.1}, with a concomitant change of line symbol. These curves have been computed with $u_*/u_{\rm th}^0=2$. (b) Stability diagram in the plane $(\beta/m,u_*/u_{\rm th}^0)$. The line separating convective from absolute instabilities corresponds to vanishing group velocity. For both panels, the other parameters are $\Lsat / H = 10$, $\Fr = 0.1$ and $H/z_0 = 10^2$.}
\label{fig:ConvectifAbsolu}
\end{figure}
%

\subsection{Convective-absolute instability transition}
\label{convectif-absolu}

We wish now to investigate the nature of this instability. In an absolute instability, some of the unstable modes propagate upstream while others propagate downstream. The instability then develops exponentially in time with a pattern that remains homogeneous in space. In a convective instability, all the unstable modes propagate in the same direction so that the instability develops in space and form a pattern stationary in time. To determine whether the bar instability is convective or absolute, the group velocity $v_g = \frac{d(ck_x)}{dk_x}$ must be analysed -- see the review of Chomaz (2005). We have computed $v_g$ for the most unstable mode ($\sigma=\sigma_{\rm max}$), as well as for the marginally stable modes ($\sigma=0$). There are two such marginally stable modes $k_+$ and $k_-$, respectively above and below $k_{\rm max}$. Note that $k_-$ vanishes for $u_*/u_{\rm th}^0 \to 1$  (Fig.~\ref{fig:lsat1}). These three values of $v_g$ are plotted in Fig.~\ref{fig:ConvectifAbsolu}a as a function of the channel aspect ratio. We can see that both $v_{\rm max}$ and $v_+$ are positive, while $v_-$ switches from positive to negative values as $\beta/m$ increases.

Positive values of the group velocity $v_-$ are associated to a convective nature: all unstable perturbations are convected downstream while amplified. Conversely, negative values of $v_-$ are associated to absolute instabilities: there exist a range of $k$ between $k_-$ and $k_{\rm max}$ for which modes are unstable but propagate upstream. We have plotted the transition between the two in Fig.~\ref{fig:ConvectifAbsolu}b in the plane $\beta/m$ \emph{vs} $u_*/u_{\rm th}^0$ (the line for which $v_-=0$). Since $k_-$ vanishes when $u_*/u_{\rm th}^0 \to 1$, the convective region shrinks in this limit: as a consequence, an absolute instability is expected very close to this transport threshold.

\subsection{From alternate to multiple bars}
\label{BarsBraiding}
 \begin{figure}
\centerline{\includegraphics{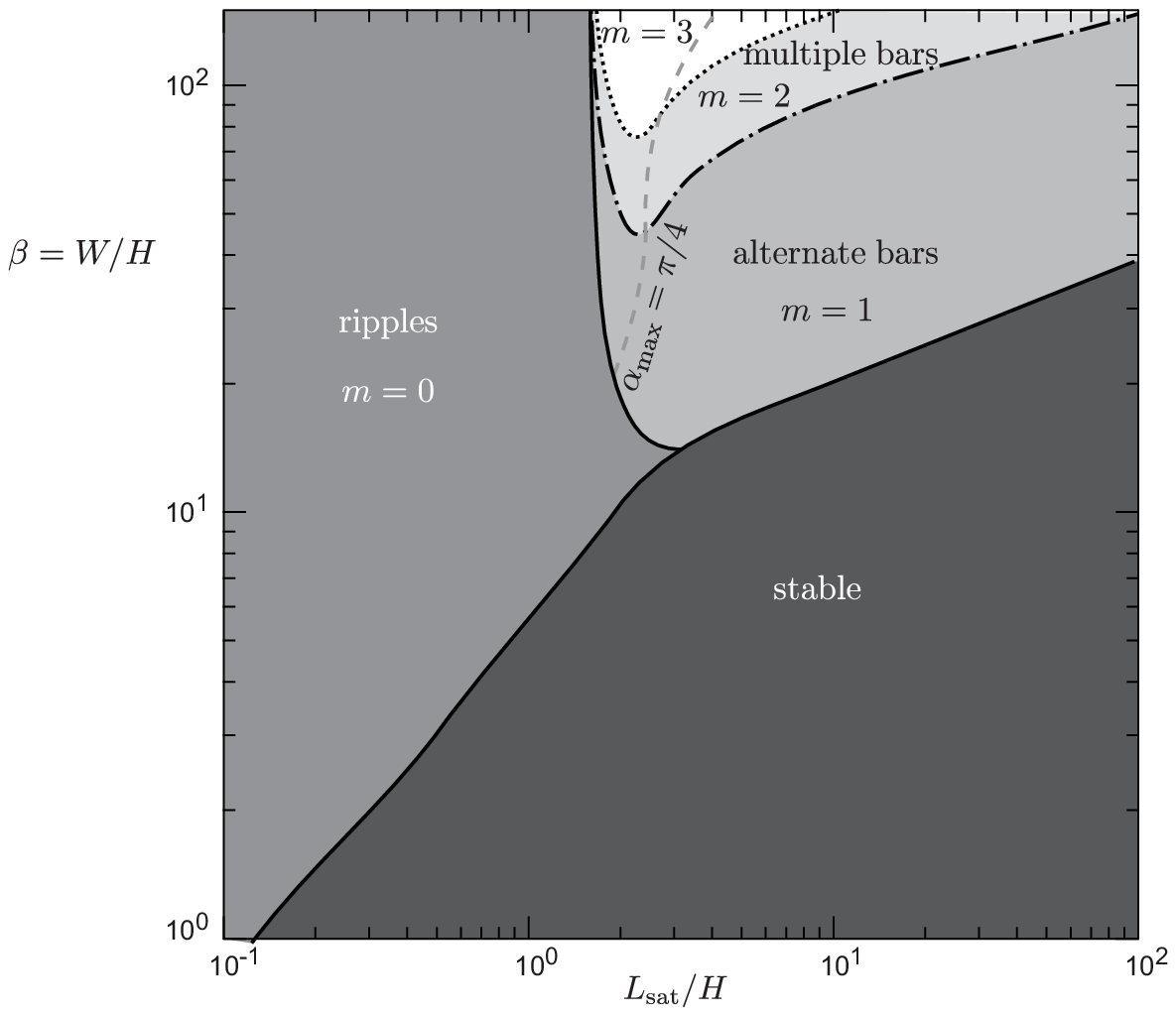}}
\caption{Diagram showing the different bedform regions in the plane ($\beta$,$\Lsat/H$). In the bottom-right dark-grey region, the bed is stable. In the left middle-grey region, ripples ($m=0$) are the most unstable bedforms. In the remaining up-right part of the plane, higher modes ($m>0$) are dominant: the dashed-dotted line separates alternate bars ($m=1$) from double bars ($m=2$). The dotted lines shows the transition to $m=3$. The dashed line ($\alpha_{\rm max} = \pi/4$) separates rather transverse to more elongated bars. These curves have been computed with $u_*/u_{\rm th}^0=2$, $\Fr = 0.1$ and $H/z_0 = 10^2$.}
\label{diagramBetaLsatsH}
\end{figure}

In paragraph~\ref{dispersion_relation}, we have seen that the presence of the banks selects a discrete though infinite set of modes. More precisely, the boundary conditions lead to a selection of the transverse wavenumber: $k_y H = \beta / m \pi$. Consider an initial sediment bed randomly disturbed, say, by a white noise. Then all modes, characterised by the pair ($k_x$,$m$) have statistically the same mean squared amplitude. The mode that emerges is the most unstable one, i.e. that with the maximum growth rate $\sigma$ with respect to $k_x$ \emph{and} $m$. By contrast, in the previous paragraphs, we have analysed the properties of the maximum growth rate at a given transverse number $m$. The phase diagram of Fig.~\ref{diagramBetaLsatsH} presents the value of $m$ selected as a function of $L_{\rm sat}/H$ and $\beta$, keeping $u_*/u_{\rm th} = 2$ fixed. 

The diagram first shows that the saturation number $L_{\rm sat}/H$ controls the transition from transverse patterns (ripples) to bars. The second dominant feature of the diagram is the existence of an instability threshold. At low $L_{\rm sat}/H$, the dimensionless parameter controlling this threshold is the ratio $W/L_{\rm sat}$. At large $L_{\rm sat}/H$, the saturation length becomes a subdominant parameter and the bedform pattern becomes controlled by the channel aspect ratio $\beta$. At small $\beta$, a flat sand bed is stable. In an intermediate range of $\beta$, alternate bars ($m=1$) are the most unstable modes. Finally, increasing the aspect ratio $\beta$, multiple bars form with an increasing value of $m$. This last regime could explain the formation and the dynamics of bars in braided rivers (Fig.~\ref{schematics}b).

\subsection{Discussion and perspectives}
\label{DiscussionPerspectives}
In conclusion, we have pointed out different artefacts in the standard theory of bar instability, which result from spurious hydrodynamical predictions of Saint-Venant shallow water equations at large wavenumbers. Amongst these artefacts, two must be particularly emphasised: the existence of a maximum growth rate in the linear instability and the stability of transverse modes. Our analysis, based instead on Reynolds averaged Navier-Stokes equations, reveals the fundamental importance of the relaxation of transport towards equilibrium: the dimensionless number characterising the influence of this dynamical mechanism is the saturation number $L_{\rm sat}/H$ turn out controls the transition from ripples (transverse patterns) to bars (inclined patterns). The qualitative aspects of the bar instability previously identified are recovered at large $L_{\rm sat}/H$. The instability threshold and the transition from alternate bars to multiple bars are mostly controlled by the river aspect ratio $\beta$, with a subdominant dependence on the saturation number. The instability presents at low wavenumber, a transition from upstream to downstream propagating modes. In particular, the so-called 'resonant conditions' where both the growth rate $\sigma$ and the propagation speed $c$ vanish are recovered. 

Our findings have important consequences for the morphology of natural rivers. Indeed, small values of $\Lsat/H$ are typical of bedload transport. In this case, the primary linear instability of a flat erodible bed leads to the emergence of transverse ripples, i.e. to the mode $m=0$. Conversely, large values of $\Lsat/H$ are typical of transport of particles in turbulent suspension (Claudin et al., 2011). Then the linear stability analysis predicts the emergence of alternate bars with an angle between $\pi/4$ and $\pi/2$. In conclusion, the transition from bed load to suspended load in sandy rivers is associated to a sharp change of bedforms. While centimetre-scale sand ripples form when bedload dominates, they are erased during floods and alternate bars then form.

To go further in the quantitative interpretation of natural bedforms a calibration effort is needed, and the saturation number $\Lsat/H$ should be systematically measured in-situ and in flumes, besides standard parameters. Furthermore, as the range of unstable modes is large, pattern coarsening --~amongst other possible non linear effects~-- is expected to take place. As a consequence, the final wavelength measured in natural rivers or flumes may be substantially different from that predicted by the linear instability analysis. Finally, it would be interesting to predict the evolution of the multiple bars composing braided streams, using the modal analysis developed here, in order to test quantitatively the relevance of the approach.

We also foresee three promising theoretical perspectives on this subject. First, it would be interesting to derive a low dimensional model for hydrodynamics  that would not present the artefacts of standard Saint-Venant equations (Ruyer-Quil \& Manneville, 1998, 2000; Luchini \& Charru, 2010). Second, the proper asymptotic matching theory, including the boundary layers close to the banks, remains to be performed. Last, the current theory of river meandering is based on the hypothesis of weak coupling between bank distortions and bedforms, with a strong emphasis on the `resonant' conditions. It would be interesting to perform a complete linear stability analysis of a channel, including erosion and deposition in the bank region, based of RANS hydrodynamical equations.

\vspace*{0.3cm}
\noindent
\rule[0.1cm]{3cm}{1pt}

We thank S. Rodrigues for the permission to use his photograph of alternate bars in the Loire river. This work has benefited from the financial support of the Agence Nationale de la Recherche, grant `Zephyr' ($\#$ERCS07\underline{\ }18).



\begin{thebibliography}{}

\bibitem[Andreotti et al. (2012)]{ACDDF12}
Andreotti, B., Claudin, P., Devauchelle, O. Dur\'an, O. \& Fourri\`ere, A. 2012 Bedforms in a turbulent stream: ripples, chevrons and anti-dunes. \textit{J. Fluid Mech.} \textbf{690}, 94-128.

\bibitem[Andreotti et al. (2002)]{ACD02}
Andreotti, B., Claudin, P. \& Douady, S. 2002 Selection of dune shapes and velocities. Part 2: A two-dimensional modelling. \textit{Eur. Phys. J. B} \textbf{28}, 341-352.

\bibitem[Ayotte et al. (1994)]{AXT94}
Ayotte, K. W., Xu, D. \& Taylor, P. A. 1994 The impact of turbulence closure schemes on predictions of the mixed spectral finite-difference model for flow over topography. \textit{Boundary-Layer Met.} \textbf{68}, 1-33.

\bibitem[Blondeaux \& Seminara (1985)]{BS85}
Blondeaux, P. \& Seminara, G. 1985 A unified bar-bend theory of river meanders. \textit{J. Fluid Mech.} \textbf{157}, 449-470.

\bibitem[Callander (1969)]{C69}
Callander, R.A. 1969 Instability and river channels. \textit{J. Fluid Mech.} \textbf{36}, 465-480.

\bibitem[Chang \& Simons (1970)]{CS70}
Chang, H.Y. \& Simons, D.B. 1970 The bed configuration of straight sand-bed channels when flow is nearly critical. \textit{J. Fluid Mech.} \textbf{42}, 491-495.

\bibitem[Chang et al. (1971)]{CSW71}
Chang, H.Y., Simons, D.B. \& Woolhiser, D.A. 1971 Flume experiments on alternate bar formation. \textit{J. Waterways Harbors Costal Eng. Div.} \textbf{97}, 155-165.

\bibitem[Charru (2006)]{C06}
Charru, F. 2006 Selection of the ripple length on a granular bed. \textit{Phys. Fluids} {\bf 18}, 121508.

\bibitem[Charru et al. (2013)]{CAC13}
Charru, F., Andreotti, B. \& Claudin, P. 2013 Sand ripples and dunes. \textit{Annu. Rev. Fluid Mech.}, accepted.

\bibitem[Chomaz (2005)]{C05}
Chomaz, J.-M. 2005 Global instabilities in spatially developping flows: non-normality and nonlinearity. \textit{Annu. Rev. Fluid Mech.} \textbf{37}, 357-392.

\bibitem[Claudin \& Andreotti (2006)]{CA06}
Claudin, P. \& Andreotti, B. 2006 A scaling law for aeolian dunes on Mars, Venus, Earth, and for sub-aqueous ripples. \textit{Earth Pla. Sci. Lett.} {\bf 252}, 30-44.

\bibitem[Claudin et al. (2011)]{CCA11}
Claudin, P., Charru, F. \& Andreotti, B. 2011 Transport relaxation time and length scales in turbulent suspensions. \textit{J. Fluid Mech.} \textbf{671}, 491-506.

\bibitem[Claudin et al. (2012)]{CWA12}
Claudin, P., Wiggs, G.F.S. \& Andreotti, B. 2012 Field evidence for the upwind velocity shift at the crest of low dunes. Submitted to Boundary-Layer Meteorology, \texttt{ArXiv:1205.4411}.

\bibitem[Colombini et al. (1987)]{CST87}
Colombini, M. Seminara, G. \& Tubino, M. 1987 Finite-Amplitude Alternate Bars. \textit{J. Fluid Mech.} \textbf{181}, 213-232.

\bibitem[Crosato \& Musselman(2009)]{CM09}
Crosato, A., \& Mosselman, E. 2009 Simple physics-based predictor for the number of river bars and the transition between meandering and braiding. \textit{Water Ressources Res.} \textbf{45}, W03424.

\bibitem[Crosato et al. (2011)]{CMDU11}
Crosato, A., Mosselman, E., Desta, F.B. \& Uijttewaal, W.S.J. 2011 Experimental and numerical evidence for intrinsic nonmigrating bars in alluvial channels. \textit{Water Ressources Res.} \textbf{47}, W03511.

\bibitem[Devauchelle et al. 2010a]{DMLJLM10}
Devauchelle, O., Malverti, L., Lajeunesse, E., Josserand, C., Lagr\'ee, P.-Y., \& M\'etivier, F. 2010 Rhomboid beach pattern: a laboratory investigation. \textit{J. Geophys. Res.} \textbf{115}, F02017.

\bibitem[Devauchelle et al. (2010b)]{DMLLJN10}
Devauchelle, O., Malverti, L., Lajeunesse, E., Lagr\'ee, P.-Y., Josserand, C. \& Nguyen Thu-Lam, K.-D. 2010 Stability of bedforms in laminar flows with free surface: from bars to ripples. \textit{J. Fluid Mech.} \textbf{642}, 329-348.

\bibitem[Dey (2003)]{D03}
Dey, S., 2003 Threshold of sediment motion on combined transverse and longitudinal sloping beds. \textit{J. Hydraul. Res.} \textbf{41}, 405-415.

\bibitem[{Dur\'an et al. (2011)}]{DCA11}
Dur\'an, O., Claudin, P. \& Andreotti B.  2011 On aeolian transport: grain-scale interactions, dynamical mechanisms and scaling laws. Review Article. Aeolian Research \textbf{3}, 243-270.

\bibitem[{Dur\'an et al. (2012)}]{DAC12}
Dur\'an, O., Andreotti B. \& Claudin, P. 2012 Numerical simulation of turbulent sediment transport, from bed load to saltation. To appear in \textit{Phys. Fluids}, \texttt{ArXiv:1111.6898}. 

\bibitem[Einstein (1950)]{E50}
Einstein, H.A. 1950 The bed load function for sedimentation in open channel flows. \textit{Technical bulletin} (US Dept. of Agriculture), \textbf{1026}, 1-69.

\bibitem[Elbelrhiti et al. 2005]{ECA05}
Elbelrhiti, H., Claudin, C. \& Andreotti, B. 2005 Field evidence for surface wave induced instability of sand dunes. \textit{Nature} {\bf 437}, 720-723.

\bibitem[Engelund \& Skovgaard (1973)]{ES73}
Engelund, F. \& Skovgaard, O. 1973 On the origin of meandering and braiding in alluvial streams. \textit{J. Fluid Mech.} \textbf{57}, 289-302.

\bibitem[Fernandez Luque \& van Beek (1976)]{LvB76}
Fernandez Luque, R. \& van Beek, R. 1976 Erosion and transport of bed-load sediment. \textit{J. Hydraul. Res.} \textbf{14}, 127-144.

\bibitem[Fourri\`ere et al. (2010)]{FCA10}
Fourri\`ere, A., Claudin P. \& Andreotti, B. 2010 Bedforms in a turbulent stream: formation of ripples by primary linear instability and of dunes by non-linear pattern coarsening. \textit{J. Fluid Mech.} \textbf{649}, 287-328.

\bibitem[Freds\o e (1978)]{F78}
Freds\o e, J. 1978 Meandering and braiding of rivers. \textit{J. Fluid Mech.} \textbf{84}, 609-624.

\bibitem[Fujita \& Muramoto (1982)]{FM82}
Fujita, Y. \& Muramoto, Y. 1982 Experimental study on stream channel processes in alluvial rivers. \textit{Bull. Disas. Prev. Res. Inst.} (Kyoto Univ.) \textbf{32}, 49-96.

\bibitem[Fujita \& Muramoto (1985)]{FM85}
Fujita, Y. \& Muramoto, Y. 1985 Studies on the process of development of alternate bars. \textit{Bull. Disas. Prev. Res. Inst.} (Kyoto Univ.) \textbf{35}, 55-86.

\bibitem[Fukuoka (1989)]{F89}
Fukuoka, S., 1989 Finite amplitude development of alternate bars. 
in \emph{River Meandering}, edited by S.Ikeda and G.Parker, Water Res. Monogr. \textbf{12}, AGU, Washington, D.C., 237-265.

\bibitem[Garc\'\i a \& Ni\~no (1993)]{GN93}
Garc\'\i a, M. \& Ni\~no, Y. 1993 Dynamics of sediment bars in straight and meandering channels: experiments on the resonance phenomenon. \textit{J. Hydraul. Res.} \textbf{31}, 739-761.

\bibitem[Ikeda (1983)]{I83}
Ikeda, H. 1983 Experiments on bedload transport, bedforms, and sedimentary structures using fine gravel in the 4-meter-wide flume. \textit{Environ. Res. Center pap.} (Tsukuba Univ.) \textbf{2}, 1-78.

\bibitem[Jackson \& Hunt (1975)]{JH75}
Jackson, P. S. \& Hunt, J. C. R. 1975 Turbulent wind flow over a low hill. \textit{Q. J. R. Meteorol. Soc.} \textbf{101}, 929-955.

\bibitem[Julien (1998)]{J98}
Julien, P.Y. 1998 Erosion and sedimentation. \textit{Cambridge University Press}.

\bibitem[Kennedy (1963)]{K63}
Kennedy, J.F. 1963 The mechanics of dunes and antidunes in erodible bed channels. \textit{J. Fluid Mech.} \textbf{16}, 521-544.

\bibitem[Lagr\'ee (2003)]{L03}
Lagr\'ee, P.-Y. 2003 A triple deck model of ripple formation and evolution. \textit{Phys. Fluids} \textbf{15}, 2355.

\bibitem[Lajeunesse et al. (2010)]{LMC10} 
Lajeunesse, E., Malverti, L. \& Charru, F. 2010 Bedload transport in turbulent flow at the grain scale: experiments and modeling. \textit{J. Geophys. Res.} \textbf{115}, F04001.

\bibitem[Lanzoni (2000a)]{L00a}
Lanzoni, S. 2000 Experiments on bar formation in a straight flume. 1. Graded sediment. \textit{Water Resources Res.} \textbf{36}, 3337-3349.

\bibitem[Lanzoni (2000b)]{L00b}
Lanzoni, S. 2000 Experiments on bar formation in a straight flume. 2. Uniform sediment. \textit{Water Resources Res.} \textbf{36}, 3351-3363.

\bibitem[Lisle et al. (1991)]{LII91}
Lisle, T.E., Ikeda, H. \& Iseya, F. 1991 Formation of stationary alternate bars in a steep channel with mixed-size sediment: a flume experiment. \textit{Earth Surf. Proc. Landforms} \textbf{16}, 463-469.

\bibitem[Lisle et al. (1997)]{LPIIK97}
Lisle, T.E., Pizzuto, J.E., Ikeda, H., Iseya, F. \& Kodama, Y. 1997 Evolution of a sediment wave in an experimental channel. \textit{Water Resources Res.} \textbf{33}, 1971-1981.

\bibitem[Luchini \& Charru (2010)]{LC10}
Luchini, P. \& Charru, F. 2010 Consistent section-averaged equations of quasi-one-dimensional laminar flow. \textit{J. Fluid Mech.} \textbf{656}, 337-341.

\bibitem[Meyer-Peter \& M\"uller (1948)]{MM48}
Meyer-Peter, E. \& M\"uller, R. 1948 Formulas for bed load transport. Proc., 2nd Meeting, IAHR, Stockholm, Sweden, 39-64.

\bibitem[Parker(1976)]{P76}
Parker, G. 1976 On the cause of characteristic scales of meandering and braiding in rivers. \textit{J. Fluid Mech.} \textbf{76}, 457-480.

\bibitem[Poggi et al. (2007)]{PKAR07}
Poggi, D. , Katul, G. G., Albertson, J. D. \& Ridolfi, L. 2007 An experimental investigation of turbulent flows over a hilly surface. \textit{Phys. Fluids} \textbf{19}, 036601.

\bibitem[Ruyer-Quil \& Manneville (1998)]{RM98}
Ruyer-Quil, C. \& Manneville, P. 1998 Modeling film flows down inclined planes. \textit{Eur. Phys. J. B} \textbf{6}, 277-292.

\bibitem[Ruyer-Quil \& Manneville (2000)]{RM00}
Ruyer-Quil, C. \& Manneville, P. 2000 Improved modeling film flows down inlined planes. \textit{Eur. Phys. J. B} \textbf{15}, 357-369.

\bibitem[Schielen et al. (1993)]{SDS93}
Schielen, R., Doelman, A. \& de Swart, H.E. 1993 On the nonlinear dynamics of free bars in straight channels. \textit{J. Fluid Mech.} \textbf{252}, 325-356.

\bibitem[Schumm \& Khan (1972)]{SK72}
Schumm, S.A. \& Khan, H.R. 1972 Experimental study of channel patterns. \textit{Geol. Soc. Am. Bull.} \textbf{83}, 1755-1770.

\bibitem[Seminara (2010)]{S10}
Seminara, G. 2010 Fluvial sedimentary patterns. \textit{Annu. Rev. Fluid Mech.} \textbf{42}, 43-66.

\bibitem[Seminara \& Tubino (1989)]{ST89}
Seminara, G. \& Tubino, M., 1989 Alternate bars and meandering: free, forced a mixed interactions. 
in \emph{River Meandering}, edited by S.Ikeda and G.Parker, Water Res. Monogr. \textbf{12}, AGU, Washington, D.C., 267-320.

\bibitem[Tubino et al. (1999)]{TRZ99}
Tubino, M., Repetto, R. \& Zolezzi, G. 1999 Free bars in rivers. \textit{J. Hydraul. Res.} \textbf{37}, 759-775.

\bibitem[Tubino \& Seminara (1990)]{TS90}
Tubino, M., Seminara, G., 1990. Free-forced interactions in developing meanders and suppression of free bars. \textit{J. Fluid Mech.} \textbf{214}, 131-159.

\bibitem[Valance 2005]{V05}
Valance, A. 2005 Formation of ripples over a sand bed submitted to a turbulent shear flow. \textit{Eur. Phys. J. B} \textbf{45}, 433-442.

\bibitem[Valance \& Langlois 2005]{VL05}
Valance, A. \& Langlois, V. 2005 Ripple formation over a sand bed submitted to a laminar shear flow. \textit{Eur. Phys. J. B} \textbf{43}, 283-294.

\end{thebibliography}
\end{document}